\documentclass[12pt]{article} 
\usepackage[letterpaper, margin = 1in]{geometry}
\usepackage[utf8]{inputenc}
\usepackage{graphicx}
\usepackage{amsmath}
\usepackage{multirow}
\usepackage{siunitx}
\usepackage{textcomp}
\usepackage[dvipsnames]{xcolor}
\usepackage{authblk}
\usepackage{upgreek}
\usepackage[labelfont=bf]{caption}
\usepackage[super,comma,numbers]{natbib} 
\usepackage{setspace} 

\usepackage{caption}

\def\X{{\bf X}}
\def\T{{\bf T}}
\def\t{{\bf t}}
\def\R{{\bf R}}
\def\r{{\bf r}}
\def\p{{\bf p}}
\def\s{{\bf s}}
\def\e{{\bf e}}
\def\y{{\bf y}}
\def\yhat{{\bf \hat{y}}}
\def\P{{\bf P}}
\def\Proj{\bf{P}^{\perp}}
\def\Projt{\bf{P}^{\perp'}}
\def\q{{\bf q}}
\def\VIP{\text{VIP}}
\def\Ltest{\mathrm{L_{test}}}
\def\TP{\text{TP}}
\def\FP{\text{FP}}

\def\P{\text{P}}
\def\N{\text{N}}
\def\a{{\bf a}}
\def\b{{\bf b}}
\def\c{{\bf c}}
\def\d{{\bf d}}
\def\e{{\bf e}}
\def\f{{\bf f}}
\def\g{{\bf g}}
\DeclareUnicodeCharacter{2212}{-}

\begin{document}

\singlespacing
\setstretch{1.5}

\title{\vspace{-3em}\Large \textbf{Breath analysis by ultra-sensitive broadband laser spectroscopy detects SARS-CoV-2 infection}}
\author[1,2,*]{\normalsize Qizhong Liang}
\author[1,3]{Ya-Chu Chan}
\author[1,2,8]{Jutta Toscano}
\author[4]{Kristen K. Bjorkman}
\author[4,5]{Leslie A. Leinwand}
\author[4,6]{Roy Parker}
\author[7]{Eva S. Nozik}
\author[1,2,3]{David J. Nesbitt}
\author[1,2,*]{Jun Ye}

\affil[1]{\scriptsize JILA, National Institute of Standards and Technology and University of Colorado, Boulder, CO 80309}
\affil[2]{Department of Physics, University of Colorado, Boulder, CO 80309}
\affil[3]{Department of Chemistry, University of Colorado, Boulder, CO 80309}
\affil[4]{BioFrontiers Institute, University of Colorado, Boulder, CO 80303}
\affil[5]{Department of Molecular, Cellular and Developmental Biology, University of Colorado, Boulder, CO 80303}
\affil[6]{Department of Biochemistry and HHMI, University of Colorado, Boulder, CO 80303}
\affil[7]{Cardiovascular Pulmonary Research Laboratories, Departments of Pediatrics and Medicine, and Division of Pediatric Critical Care Medicine, University of Colorado Anschutz Medical Campus, Aurora, CO 80045}
\affil[8]{Present address: Department of Chemistry, University of Basel, Klingelbergstrasse 80, 4056 Basel, Switzerland}
\affil[*]{Corresponding authors: Qizhong.Liang@colorado.edu, Ye@jila.colorado.edu}

\date{}
\maketitle
\thispagestyle{empty}

\doublespacing
\small

\pagenumbering{arabic}

\normalsize
\singlespacing
\setstretch{1.5}

\textbf{Rapid testing is essential to fighting pandemics such as COVID-19, the disease caused by the SARS-CoV-2 virus. Exhaled human breath contains multiple volatile molecules providing powerful potential for non-invasive diagnosis of diverse medical conditions. We investigated breath detection of SARS-CoV-2 infection using cavity-enhanced direct frequency comb spectroscopy (CE-DFCS), a state-of-the-art laser spectroscopic technique capable of a real-time massive collection of broadband molecular absorption features at ro-vibrational quantum state resolution and at parts-per-trillion volume detection sensitivity. Using a total of 170 individual breath samples (83 positive and 87 negative with SARS-CoV-2 based on Reverse Transcription Polymerase Chain Reaction tests), we report excellent discrimination capability for SARS-CoV-2 infection with an area under the Receiver-Operating-Characteristics curve of 0.849(4). Our results support the development of CE-DFCS as an alternative, rapid, non-invasive test for COVID-19 and highlight its remarkable potential for optical diagnoses of diverse biological conditions and disease states.}

\clearpage
\doublespacing
\section{Introduction}

The difficulty to rapidly and accurately detect Severe Acute Respiratory Syndrome Coronavirus 2 (SARS-CoV-2) infection has been a barrier to the response throughout the Coronavirus Disease 2019 (COVID-19) pandemic~\cite{Ritchie_COVIDDeaths}. The current gold standard method, Reverse Transcription Polymerase Chain Reaction (RT-PCR) test to detect viral RNA~\cite{CDC_RTPCR}, requires appropriate sample collection and storage for accuracy, and is time-consuming~\cite{Mei_PCRtakeslong}. Sampling is typically invasive (e.g., nasal swab), contributing to test hesitancy. The real-time assessment of community prevalence, implementation of public health protocols, and timely anti-viral intervention for high-risk people~\cite{Dan_Turnaround,Saravolatz_paxlovid}, would all benefit significantly from the development of rapid, safe, sensitive, and non-invasive detection methods for SARS-CoV-2 infection, particularly with recent variants showing an increased epidemic growth rate~\cite{Backer_SerialInterval}.

Exhaled breath analysis is an attractive alternative to RT-PCR detection of SARS-CoV-2 infection as it is non-invasive and can return real-time measurements~\cite{Wang_laserBreathReview,Arnold_BreathReview}. Early studies to develop breath-based COVID-19 diagnosis included nanomaterial-based sensors~\cite{Shan_COVID_NanoSensor,ZAMORAMENDOZA_COVID_NanoSensor}, ion-mobility spectrometry~\cite{RUSZKIEWICZ_COVID_GCIMS,Chen_COVID_GCIMS}, and mass spectrometry~\cite{Ibrahim_COVID_Massspec,GRASSINDELYLE_COVID_PTRMassSpec}. A COVID-19 breath diagnostic test based on Gas Chromatography-Mass Spectrometry (GC-MS) was recently granted emergency use authorization by the U.S. Food and Drug Administration after its validation with over 2,409 individuals, reporting \SI{91.2}{\%} sensitivity and \SI{99.3}{\%} specificity~\cite{FDA_FirstCOVTest,FDA_InspectIRreport}. While GC-MS currently represents one of the most powerful techniques for breath analysis due to its superior detection sensitivity and specificity~\cite{Wang_laserBreathReview,Smith_RealTimeMassSpec}, breath molecules present with identical mass-to-charge ratio imposes real analytical challenges for mass spectrometry to discriminate. In addition, unavoidable alteration to breath components via purification, derivatization, and thermal degradation introduced from the use of a pre-concentrator~\cite{FDA_InspectIRreport} and a high-temperature thermal process~\cite{Fang_GCMSThermalDegrade} can also hinder accurate measurement of breath profiles.

The recently-developed laser spectroscopy-based technique, namely the Cavity-enhanced Direct Frequency Comb Spectroscopy (CE-DFCS)~\cite{Liang-PNAS,Thorpe_breathalyzer}, can help overcome the analytical challenges of mass spectrometry. CE-DFCS rapidly detects and identifies molecules in exhaled breath by ultra-sensitively measuring their structure-specific absorption signals via laser light at numerous optical frequencies. It requires no sample heating or purifying and ensures chemistry-free determinations of breath profiles. Together with the superior parts-per-trillion detection sensitivity~\cite{Liang-PNAS}, and with robust specificity to discriminate between different isomeric, isobaric, and isotopologue compounds~\cite{Kranenburg_GCMSonISOMER}, this technique offers rapid, accurate, and robust information that can add to diagnosis and mechanistic insight. Recent proof-of-principle studies have demonstrated the use of CE-DFCS to monitor changes in exhaled breath profiles upon fruit intake~\cite{Liang-PNAS} and smoking~\cite{Thorpe_breathalyzer}, showing potential utility for disease diagnostics. To test if this powerful methodology may be useful for non-invasive medical diagnostics, a trial study was carried out for the first time to test its ability to identify SARS-CoV-2 infection in a young, highly vaccinated cohort as a case study.

\section{Method}

\subsection*{Human subjects}
This study was approved by the Institutional Review Board (protocol no. 21-0088) of the University of Colorado Boulder. From May 2021 to January 2022, breath samples from a total of 170 research subjects were collected with a class distribution for SARS-CoV-2 infection of 83 positives (\SI{48.8}{\%}) and 87 negatives (\SI{51.2}{\%}). Research subjects were all University of Colorado Boulder affiliates, at least 18 years old, and recruited after taking a university-provided saliva-based or nasal swab COVID-19 RT-PCR test. The general campus population was $>$\SI{90}{\%} vaccinated. No participants were severely ill or requiring hospitalization at the time of their sample collection. After receiving their COVID-19 test results, potential subjects received a study recruitment email and were asked to contact the research team within 24 hours if interested in participation. They then reviewed and signed an informed consent form, completed a questionnaire, and scheduled an appointment for the collection of their breath samples. The questionnaire collected self-reported information on sex, age, and race as well as other factors that could impact breath analysis including smoking, alcohol use, and underlying gastrointestinal symptoms. Additional information was collected on acute symptoms experienced by the positive participants. No viral genomes were sequenced, but the Colorado statewide data~\cite{CDPHE_variant} over our subject recruitment period indicates infection with several viral variants associated with several infection waves (namely, alpha, delta, and omicron) in the community. All data (i.e., informed consent form, questionnaire, and Tedlar bag ID) were collected and managed using the REDCap electronic data capture tool~\cite{Harris_REDCap1,Harris_REDCap2} hosted by the University of Colorado Denver.

\subsection*{Breath sample collection and handling}
Standard Tedlar bags (1-liter, part no. 249-01-PP, SKC Inc.) were used to collect exhaled breath. During the sample collection appointment, research subjects were asked to hold their nose and breathe through their mouth. They were instructed to inhale to full lung capacity for 1-3 s, followed by exhaling the first half of their breath to the surroundings and the second half into the bag until the latter was above $\sim$\SI{80}{\%} full. The sample collection location was an outdoor university parking lot. The participants were not instructed to limit or control their smoking, food or alcohol intake prior to sample collection. Right after collection of one breath sample, the Tedlar bag was stored inside an air-tight container at ambient temperature and transported to the indoor lab housing the CE-DFCS setup for immediate data collection and analysis. The breath sample was warmed to \SI{37}{\degree C} for 20 mins to reduce condensation, then steadily flowed through the cleaned vacuum chamber held at room temperature (\SI{20}{\degree C}) at a rate of $\sim$1 liter per minute. Just before bag exhaustion, timely closure of the gas valves detained a portion of breath sample inside the chamber and a static pressure of \SI{50}{Torr} (\SI{67}{mbar}) was reached (without re-condensation) for spectroscopic data collection. After the measurement, the breath sample was pumped out to an exhaust line leading to the building exterior. The used Tedlar bag was autoclaved and disposed of. While direct sampling at atmospheric pressure by our breathalyzer is feasible, off-line sampling and negative pressure were adopted to ensure no SARS-CoV-2 could be introduced into the laboratory air. Spectroscopy data collection for each breath sample was completed in less than 10 minutes. This can be further reduced to about 1 second when optimized data acquisition and readout are implemented. Overall, from sample collection and transportation to completion of data analysis, the total time was less than an hour. Air samples were collected on separate days over the subject’s recruitment period at the sample collection location as control specimens.

\subsection*{CE-DFCS technique}
The working principle of the CE-DFCS breathalyzer is illustrated in Fig.~\ref{fig:Principle}a. A high-resolution broadband absorption spectrum, consisting of a total of 14,836 distinct molecular features each measured ultra-sensitively at individual optical frequencies, was recorded for each breath sample (see sample spectrum in Fig.~\ref{fig:Principle}b). The breath spectrum was processed by machine learning analysis for binary response classifications. For additional instrument details, see Ref.~\cite{Liang-PNAS}.

\subsection*{Machine learning analysis}
We employed two spectral pre-processing techniques for machine learning analysis: 1) a pattern-based approach that directly used all 14,836 molecular absorption features as the predictor variables; 2) a molecule-based approach that used 16 known small molecule compounds (H$_2$O, HDO, $^{12}$CH$_4$, $^{13}$CH$_4$, OCS, C$_2$H$_4$, CS$_2$, H$_2$CO, NH$_3$, CH$_3$OH, O$_3$, N$_2$O, NO$_2$, SO$_3$, HCl, and C$_2$H$_6$) fitted to the spectra as predictor variables. The former approach identifies all stable patterns that can be used for diagnostics, whereas the latter identifies only the patterns that can be reduced to known molecular identities, which may result in loss of utilizable chemical information but allows better interpretability into the model details. The 16 compounds were chosen due to their availability from the High-Resolution Transmission Molecular Absorption (HITRAN) database~\cite{GORDON_HITRAN2016}. While more molecules can potentially be uncovered and fitted, quantitative extraction of their identities requires cross-sectional data at our experimental conditions (\SI{20}{\degree C} temperature and \SI{50}{Torr} pressure) to be available. Unfitted species are hence not used in the molecule-based analysis despite being potentially useful to facilitate better predictive power.

To enable binary class assignment, we used Partial Least Squares-Discriminant Analysis (PLS-DA)~\cite{Chuen_PLSDA-review}. This method allows for the reduction of high-dimensionality data into a one-dimensional scalar number to differentiate between the opposing response classes (positive vs. negative). Variable Importance in the Projection (VIP) scores~\cite{CHONG_VIP} were determined for assessing the relative importance of each predictor variable. To assess predictive power, the complete dataset ($N$ = 170) was randomly divided into a training set ($n$ = 140) and the remaining as a testing set ($n$ = 30). Both sets shared the same binary class distributions as the complete data set. The training set was used for model construction (a total of 15 PLS components were constructed) and the testing set was used for a blind test to obtain a Receiver-Operating-Characteristic (ROC) curve, from which the area under the curve (AUC) value was calculated. Depending on how the complete set was divided, the AUC value obtained can vary to a certain extent. To ensure convergence, we repeated the whole process (i.e., cross-validation) for a total of 10,000 times, and each time a new training set and testing set were randomly re-selected for a new AUC value to be calculated. The ROC curves generated from the total of 10,000 cross-validation runs were averaged together to obtain an averaged ROC curve. The AUC of the averaged curve thus represents the average AUC from all cross-validation runs. To determine the AUC uncertainty, we used different training/testing partition ratios and different numbers of PLS components. All analysis code was written using MATLAB, and the PLS-DA was performed using the built-in package based on the SIMPLS algorithm~\cite{DEJONG_SIMPLS}. The supplementary file contains additional details on PLS-DA and VIP score principles, ROC averaging, and AUC uncertainties.

\section{Results}

\subsection*{Subject characteristics}
\begin{figure*} [htbp]
\centering
\includegraphics[width=0.5\textwidth]{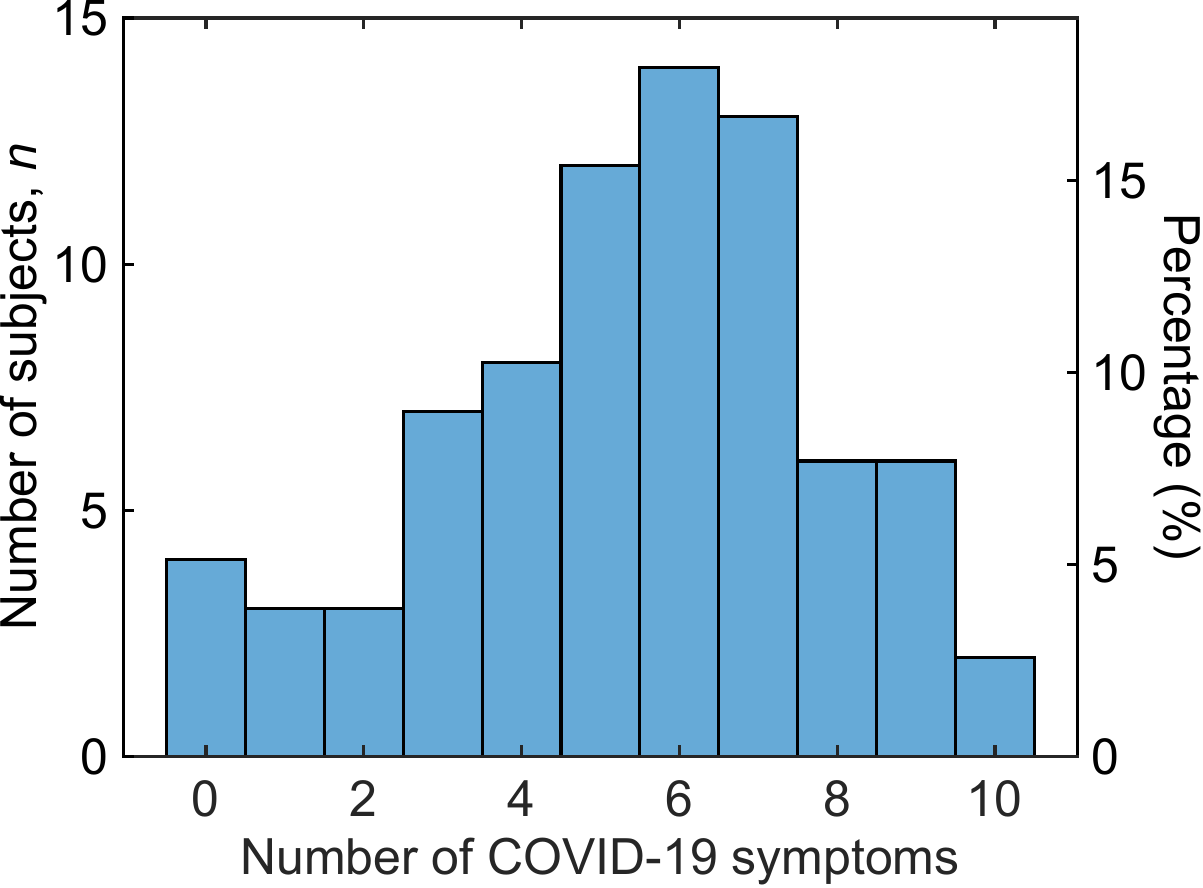}
\caption{\textbf{Number of COVID-19 symptoms experienced by the positive participants.} See Table~\ref{tab:Positive} for a list of COVID-19 symptoms. Only SARS-CoV-2 positive participants with non-missing questionnaire responses were included.}
\label{fig:SumSymp}
\end{figure*}

170 participants enrolled in this study, with characteristics summarized in Table~\ref{tab:Charac}. These included 83 (\SI{48.9}{\%}) SARS-CoV-2 positive subjects and 87 (\SI{51.2}{\%}) SARS-CoV-2 negative subjects based on prior RT-PCR tests. The median age was 22 years in the infection-positive and 24 years in the infection-negative groups (p$<$0.05). Both infection-positive and negative groups were balanced for sex (\SI{53.0}{\%} female infection-positives, \SI{49.4}{\%} female negatives). Race and ethnicity distributions were equivalent between infection-positive and negative groups. A higher number of infection-negative subjects reported a history of rare to occasional abdominal symptoms, though there was no difference in the history of lactose intolerance or constipation between the two groups.  SARS-CoV-2-positive subjects were asked additional questions regarding COVID-19-related symptoms, if any (Table~\ref{tab:Positive}). We found most subjects reported multiple symptoms (Fig.~\ref{fig:SumSymp}). Of 78 who responded, \SI{50.0}{\%} reported 5-7 of the 11 listed symptoms, \SI{5.1}{\%} were asymptomatic, and \SI{2.6}{\%} reported 10 symptoms.

\subsection*{Comparable prediction accuracy for SARS-CoV-2 infection by RT-PCR and CE-DFCS}
\begin{figure*} [htbp]
\centering
\includegraphics[width=\textwidth]{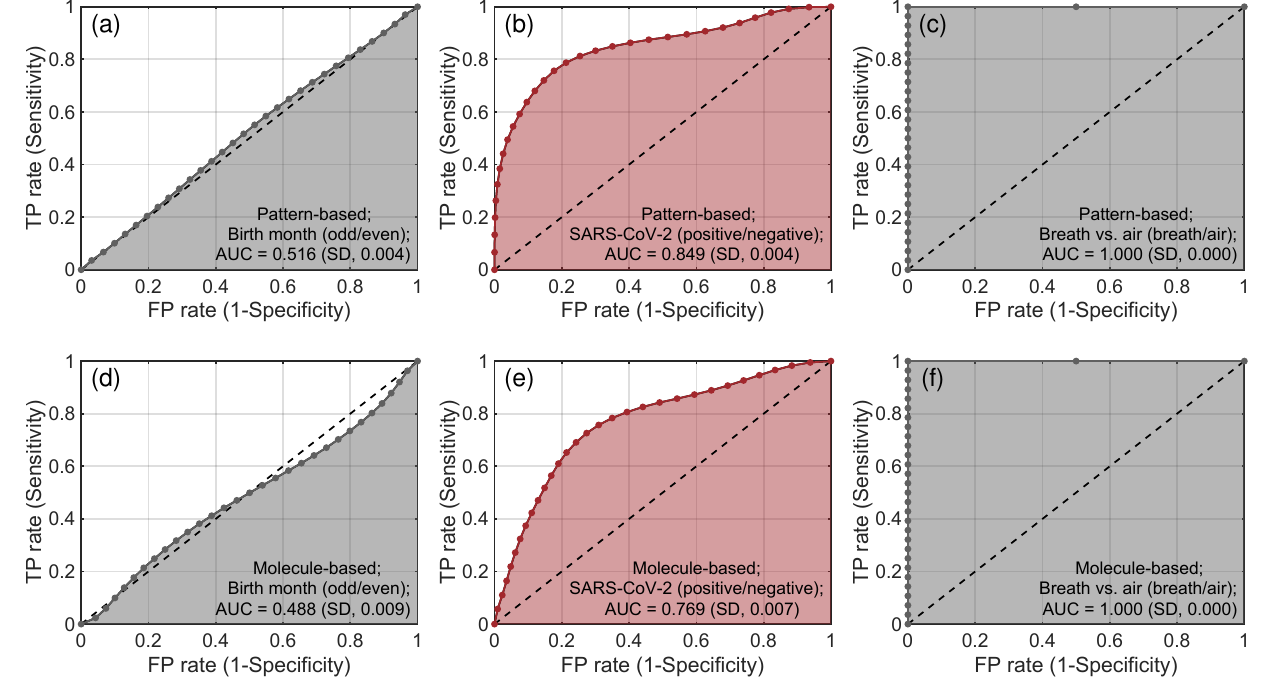} 
\caption{\textbf{Prediction performance for SARS-CoV-2 infection.} Results for SARS-CoV-2 (\b, \e) are plotted in red while two controls (\a, \c, \d, \f) for validation of the analysis methodology were plotted in black. Top panels (\a, \b, \c) and bottom panels (\d, \e, \f) show prediction results obtained by the pattern-based approach and the molecule-based approach, respectively. A control based on birth month (\a, \d) examines whether subjects were born on the even or the odd months. A control based on breath vs. ambient air (\c, \f) examines whether spectroscopy data were measured for inhaled air or exhaled breath. Obtained AUCs are reported on the figures. Respective assignment of the response classes for the two controls to positive and negative was done at random and does not carry any particular meaning. Details on cross-validations are described in the main text. TP, true positive; FP, false positive.}
\label{fig:SARS}
\end{figure*}

Breath analysis by laser spectroscopy can differentiate between SARS-CoV-2 infection positives and negatives. Using the two spectral pre-processing techniques for machine learning analysis, we found the pattern-based approach yielded an AUC of 0.849 (standard deviation [SD], 0.004) (Fig.~\ref{fig:SARS}b) and the molecule-based approach yielded an AUC of 0.769 (SD, 0.007) (Fig.~\ref{fig:SARS}e). Both approaches confirmed that significant differences in breath contents caused by SARS-CoV-2 infection was successfully detected by CE-DFCS. The classification results on SARS-CoV-2 infection should be interpreted as the co-agreement between the CE-DFCS breath test and the RT-PCR tests employed. As control experiments to validate the analysis methodology, we checked predictions for two cases with known responses: 1) a random guess based on subjects born in even vs. odd months, for which the lowest possible AUC of 0.5 is expected; 2) a perfect discrimination comparing ambient air vs. exhaled breath samples, for which one expects an AUC of 1. Both the pattern-based and molecule-based approaches confirmed expectations for results from a random sampling by birth month (Fig.~\ref{fig:SARS}a and \ref{fig:SARS}d), yielding an AUC of 0.516 (SD, 0.004) and 0.488 (SD, 0.009) respectively. With regard to ambient air vs. breath, both approaches yielded AUCs of 1.000 (SD, 0.000) (Fig.~\ref{fig:SARS}c and \ref{fig:SARS}f) and confirmed perfect discrimination criterion. These results further support the reliability of our analysis protocol. The AUC of $\sim$0.5 obtained from predictions of baseline response also suggested that our sample size was large enough to capture sufficient population diversity.

\subsection*{Pattern-based approach outperforms molecule-based approach}
\begin{figure*} [htbp]
\centering
\includegraphics[width=\textwidth]{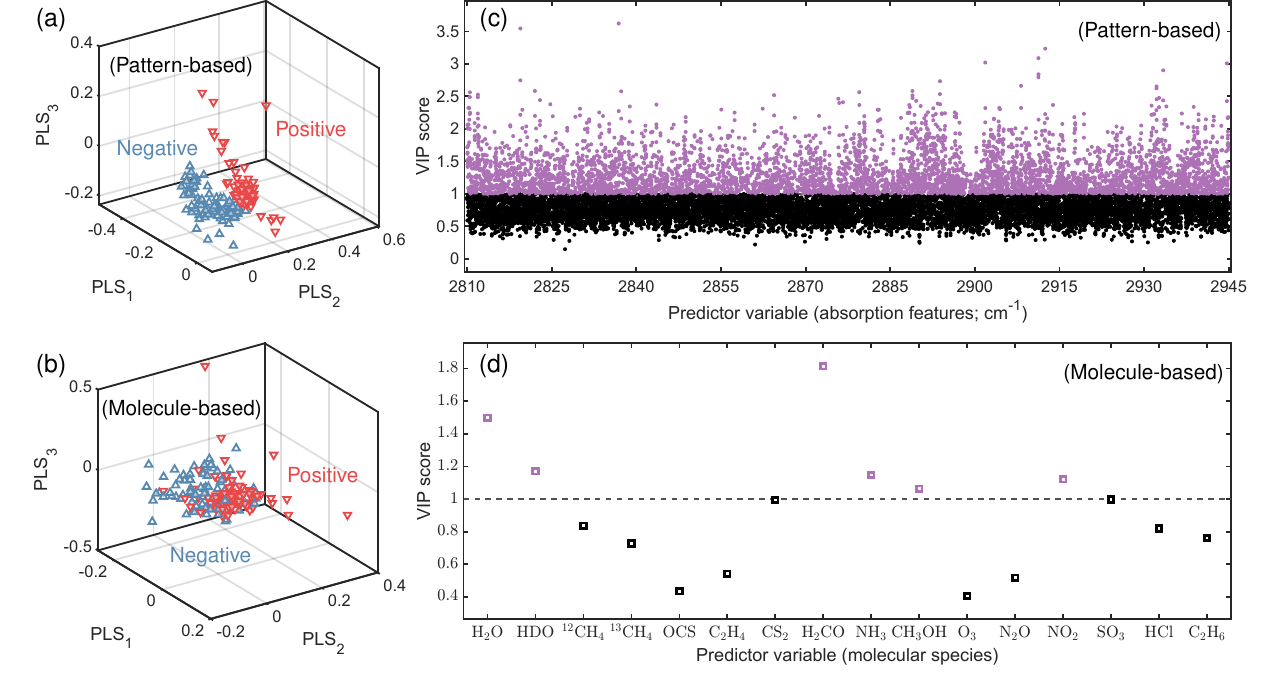}
\caption{\textbf{Pattern-based approach over molecule-based approach.} \a, \b, distribution of the subjects’ data for the first three PLS components, with red (down-pointing) and blue (up-pointing) triangles representing positive and negative research subjects, respectively. \c, \d, VIP scores showing the importance of different predictor variables in prediction making. Predictor variables with VIP scores above (or below) unity were plotted in purple (or black) and considered as important (or unimportant) for predictions. Results shown for the (\a, \c) pattern-based and (\b, \d) molecule-based approaches were calculated using the complete data set (N = 170) for SARS-CoV-2 infection.
}
\label{fig:PLS}
\end{figure*}

For SARS-CoV-2 infection, we found that the pattern-based approach clearly outperformed the molecule-based approach in prediction performance (AUC of 0.849 (SD, 0.004) vs. 0.769 (SD, 0.007)). To illustrate this result, we made use of the subjects’ distribution on the Partial Least Squares (PLS) coordinate, which allowed us to visualize which approach can better discriminate opposing response classes. We used the complete data set ($N$ = 170) for construction of the PLS coordinate space and plotted subjects' data on the first three PLS components in Figures~\ref{fig:PLS}a (pattern-based) and~\ref{fig:PLS}b (molecule-based). The results show significantly better discrimination capability was obtained by the pattern-based approach. The underperformance of the molecule-based approach could potentially be attributed to the exclusion of species with unknown identities in exhaled breath detected by CE-DFCS. As CE-DFCS acquires breath data at extremely high sensitivity, specificity, and dimensionality, applying the pattern-based approach to make full use of the wealth of chemical information collected by CE-DFCS is advantageous in that it bypasses the need for a complete molecular database to directly understand the best possible prediction power.

A notable limitation of the pattern-based approach, however, is that it does not reveal which molecules are important for making predictions, but only the optical frequencies at which they are probed. Variable importance analyzed for the pattern-based approach (Fig.~\ref{fig:PLS}c) identified prediction-important optical frequencies (VIP scores $>$ 1 and highlighted in purple) where measured absorption values were strongly discriminative between SARS-CoV-2 positives and negatives. These frequencies are distributed near-uniformly over the entire spectrum. On the other hand, variable importance analyzed for the molecule-based approach (Fig.~\ref{fig:PLS}d) identified a panel of indicative molecular species for SARS-CoV-2 infection: water (H$_2$O), semiheavy water (HDO), formaldehyde (H$_2$CO), ammonia (NH$_3$), methanol (CH$_3$OH), and nitrogen dioxide (NO$_2$). Being able to identify the molecules provides better clarity to rationalize a possible prediction. To illustrate, variable importance performed for ambient air vs. breath samples based on the molecule-based approach identified water (H$_2$O) and semiheavy water (HDO) as the only important predictor variables (data not shown). This is easy to understand because water contents were saturated in breath and hence the machine could solely rely on them for prediction. The panel of indicative molecules identified by the molecule-based approach for SARS-CoV-2 infection provides the opportunity for further studies to elucidate the pathophysiology of SARS-CoV-2 infection. 

\subsection*{Prediction performance for a list of potential confounders}
\begin{figure*} [htbp]
\centering
\includegraphics[width=\textwidth]{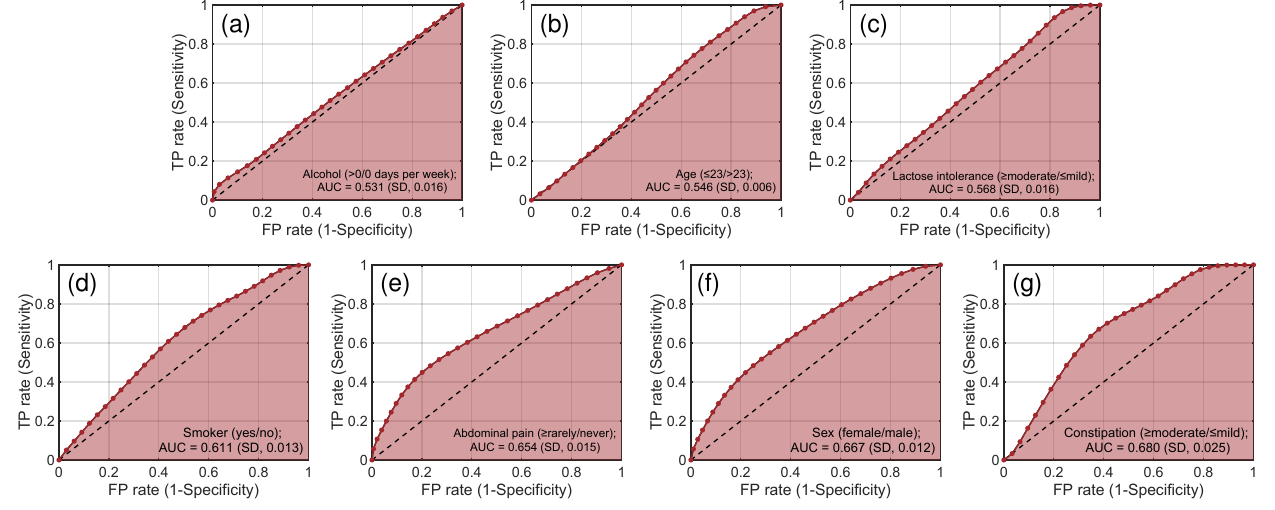} 
\caption{\textbf{Prediction performance for a list of potential counfounders.} Random guessing results (AUC $<$ 0.6) were found for (\a) alcohol use, (\b) age, and (\c) lactose intolerance. Significant differences (0.6 $\leq$ AUC $<$ 0.7) were found for (\d) smoking, (\e) abdominal pain, (\f) sex, and (\g) constipation. Class assignments for each response type are shown on the figures. For age, a median age of 23 years old was used for class assignment. All results shown were analyzed by the pattern-based approach and details on the cross-validation are described in the main text. }
\label{fig:Confounders}
\end{figure*}

We analyzed the prediction performance for a list of subject characteristics and potential factors that could confound the results. For prediction of a specific response, subjects from the complete dataset ($N$=170) were divided into opposing classes based on the self-reported questionnaire data. Results obtained using the pattern-based approach are presented in Fig.~\ref{fig:Confounders} and the group assignment criteria for different response types are listed in the panels. A summary for all prediction analyses can also be found in Table~\ref{tab:Summary}. From the results, we found random guessing predictions (AUC $<$ 0.6) for alcohol use, age, and lactose intolerance, but significant prediction capabilities for smoking, sex, abdominal pain, and constipation (0.6 $<$ AUC $<$ 0.7). On age and abdominal pain, while our subjects had modest correlations with SARS-CoV-2 infection, the significantly better predictive power for SARS-CoV-2 infection suggests that age and abdominal pain do not constitute strong confounders. The superior prediction performance for SARS-CoV-2 infection compared to the list of potential confounders analyzed could potentially be due to SARS-CoV-2 infection eliciting acute and long-term host responses caused by both virus-driven and immune system-associated factors.

\section{Discussion}

We conducted the first pilot study to evaluate the diagnostic performance of CE-DFCS. Through a case study of SARS-CoV-2 infection detection involving 170 individuals, we found our pattern-based model produced excellent mutual agreement of 0.849 (SD, 0.004) AUC between the CE-DFCS test and the RT-PCR test results. Moreover, using the molecule-based model, we identified the relative importance of different breath molecules in making predictions. Finally, we present preliminary evidence that this technique could be extended to diagnose other conditions.

Our most important finding is that breath analysis by CE-DFCS can differentiate between SARS-CoV-2 infection positives and negatives. This study builds upon our prior works in which we established the use of CE-DFCS for the characterization of exhaled breath molecular profiles upon changes in biological conditions~\cite{Liang-PNAS,Thorpe_breathalyzer}. Here, we have carried out the first trial study for CE-DFCS and employed machine learning analysis to realize robust binary diagnostics. Our study established CE-DFCS as a new diagnostic tool based on ultra-sensitive broadband laser spectroscopy. Continued assessment of CE-DFCS is important to thoroughly understand its diagnostic utility. Currently, the differences in the study designs make it difficult to compare the performance of CE-DFCS with GC-MS. The GC-MS study that has received FDA approval~\cite{FDA_FirstCOVTest,FDA_InspectIRreport} prospectively conducted RT-PCR tests and collected breath samples within 5 minutes of each other, restricted eating, drinking, or smoking for the 15 minutes preceding sample collection and excluded participation from those who had recent exposure to areas of local COVID-19 spread or close contact with COVID-19 positives. By contrast, our study had a much longer time delay from RT-PCR tests to breath sample collections (2.05 (SD, 0.95) days for the positives), and no exclusions based on travel/contact history. The time lag may result in viral clearance, and the more lenient sample collection and recruitment protocols may introduce confounders. These differences preclude a direct comparison of the two techniques. For future studies, examination of CE-DFCS’s utility in individuals with severe disease or at higher risk, such as the elderly, the unvaccinated, and those with pre-disposing co-morbidities, will be important. 

CE-DFCS may have broader applicability beyond the detection of SARS-CoV-2 infection. It may also 1) serve as a non-invasive tool for evaluation of other health or biological conditions, and 2) provide insights into disease pathogenesis. With respect to 1), our results show that CE-DFCS discriminated between subjects based on smoking history~\cite{Kushch_Smoke1,Buszewski_Smoke2}, biological sex~\cite{Buszewski_Sex1,Taylor_Sex2,Good_Sex3,Hannemann_Sex4}, as well as gastrointestinal symptoms~\cite{Pichetshote_GI1,Costello_GI2,Banik_GI3} (recurring abdominal pain and constipation). We were not able to discriminate subjects based on alcohol intake~\cite{Hlastala_alcohol1} or lactose intolerance~\cite{Costanzo_lactose1}, but this is not surprising as our subjects had not been specifically challenged with alcohol or lactose ingestion. With respect to 2), it has been recently reported~\cite{Barauna_IRspecSARSCoV2} that SARS-CoV-2 virus exhibits strong optical absorption signals within our spectral coverage (\SI{2810}{cm^{−1}}-\SI{2945}{cm^{−1}}). This signal could potentially partly originate from the C-H molecular bonds in the surface-exposed SARS-CoV-2 spike protein~\cite{Soares_SpikeProtein}. A future measurement of the viral absorption spectrum in the gas phase with proper consideration of protein structure dynamics~\cite{Lopez_ProteinAnalysis} may allow direct quantification of viral load in exhaled breath with CE-DFCS. This could allow us to examine the correlation between viral burden and other breath biomarkers and to determine the relative contributions of virus and host response to the change in breath molecular profiles. We find our results compelling enough to warrant future investigation into the applicability of CE-DFCS breath analysis to other conditions or diseases, particularly those of respiratory, gastrointestinal, or metabolic origin.

Finally, we note that ongoing rapid developments can further empower CE-DFCS in its use for medical diagnostics. Spectral range of the current CE-DFCS setup can be expanded to cover more ro-vibrational bands~\cite{Iwakuni_10um,Picque2019,Kippenberg2020,Lesko_octave}, thereby probing more discriminative features for stronger predictions. Furthermore, due to the direct measurement capability of CE-DFCS (i.e., no need for chemical treatments, pre-concentrations, and thermal processing), the technique can facilitate the creation of large-scale databases by accumulating breath data from different trial studies. This can promote the construction of deep learning model architectures~\cite{Amato_NeuralMedical,Ibrahim_IRSPEC_CHratio,Enders_IRSPEC_MolFunc} that can outperform traditional machine learning algorithms (e.g., PLS-DA) in predictive power. Recent photonics advances could potentially permit chip-scale miniaturization~\cite{Xiang_ChipComb,Jin_HighFMicroMirror,Fathy_ChipFTIR} for CE-DFCS and thus the technique could eventually be integrated into portable devices to support low-cost, widespread use and enable daily self-health monitoring on the go. 

\section{Conclusion}

We present the first trial study of laser frequency comb spectroscopy for non-invasive medical diagnostics. Our case study of SARS-CoV-2 infection detection among a total of 170 individuals finds excellent mutual agreement between CE-DFCS and RT-PCR tests and supports the development of CE-DFCS as an alternative and accurate COVID-19 test with non-invasive sampling and rapid turnaround time. While the outstanding prediction performance was achieved using the pattern-based approach, continued enrichment in the molecular absorption database will empower high-resolution comb spectroscopy to employ molecule-based approach providing comparable prediction accuracy but with significantly better model interpretability. The laser spectroscopy-based technique, capable of ultra-sensitive, multi-species, rapid and chemistry-free detection of breath molecular contents with robust isomer-, isobaric-, and isotopologue-specificity opens a complementary approach for the development of breath-based diagnostics research.

\newpage
\section*{Acknowledgements}
We thank Holly Gates-Mayer, P. B. Changala, Diego Olaya, Aaron Gilad Kusne, and Lee R. Liu for helpful discussions. This work was supported by AFOSR 9FA9550-19-1-0148, NIST, DOE DE-FG02-09ER16021, NSF CHE-2053117, NSF QLCI OMA–2016244, NSF PHY-1734006, and NIH 1R35HL139726-01. J. T. was supported by the Lindemann Trust in the form of a Postdoctoral Fellowship.

\section*{Authors contributions}
All authors contributed to the study design, results interpretation, and manuscript writing. Q. L. and Y-C. C. collected and analyzed the data. The authors declare no competing interests.

\section*{Data availability statement}
The data used for this study are available from the authors upon reasonable request.

\newpage
\bibliographystyle{osajnl}
\bibliography{Reference}

\newpage
\begingroup
  \centering
  \includegraphics[width=\linewidth]{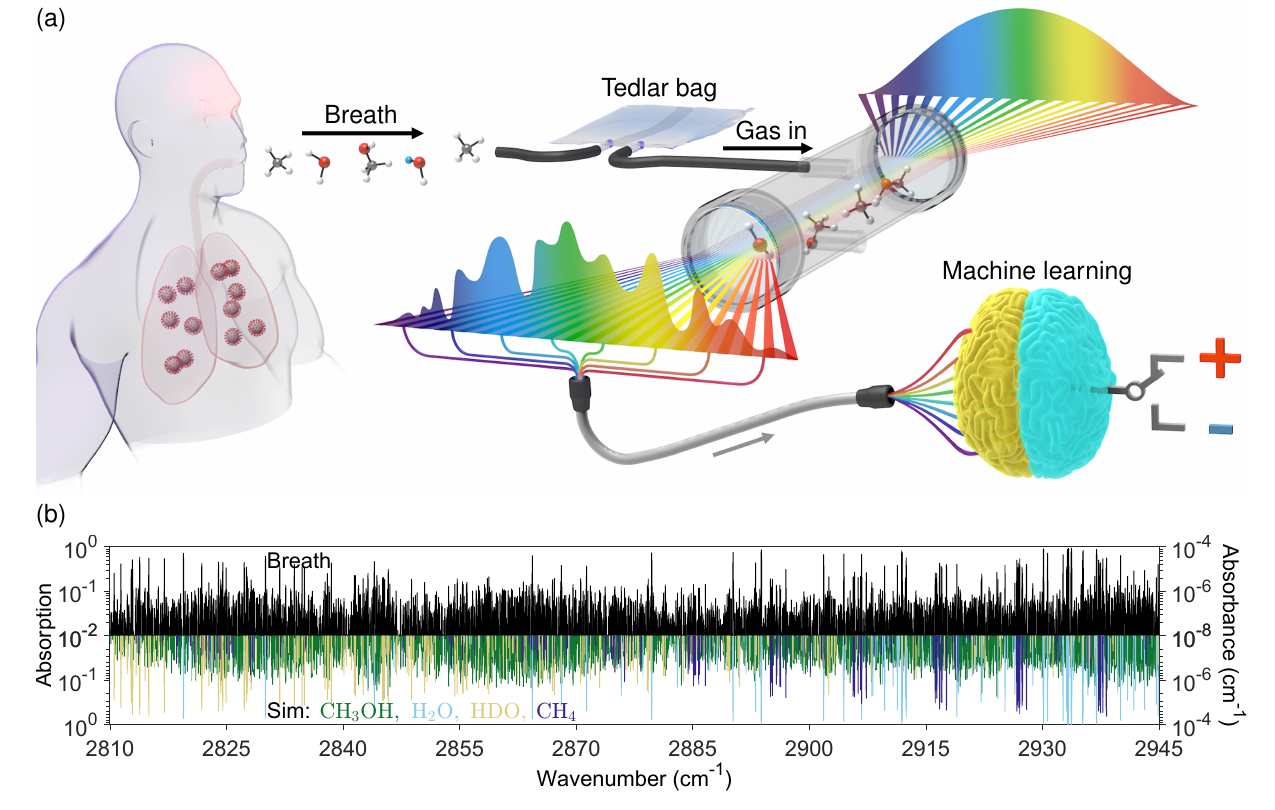}
  \captionof{figure}{\textbf{CE-DFCS breathalyzer.} \a, Schematic representation of the working principle of the device. An exhaled human breath sample was collected in a Tedlar bag and then loaded into an analysis chamber. The chamber was surrounded by a pair of high-reflectivity optical mirrors. A mid-infrared frequency comb laser interacted with the loaded sample and generated a broadband molecular absorption spectrum. The spectroscopy data was then used for supervised machine learning analysis to predict the binary response class for the research subject (either positive or negative). \b, Sample absorption spectrum collected from a research subject’s exhaled breath (black). Inverted in sign and plotted with different colors are four fitted species (CH$_3$OH, H$_2$O, HDO, and CH$_4$) that give the most dominant absorption features.}
  \label{fig:Principle}
\endgroup

\newpage
\begingroup
  \centering
  \includegraphics[width=0.95\linewidth]{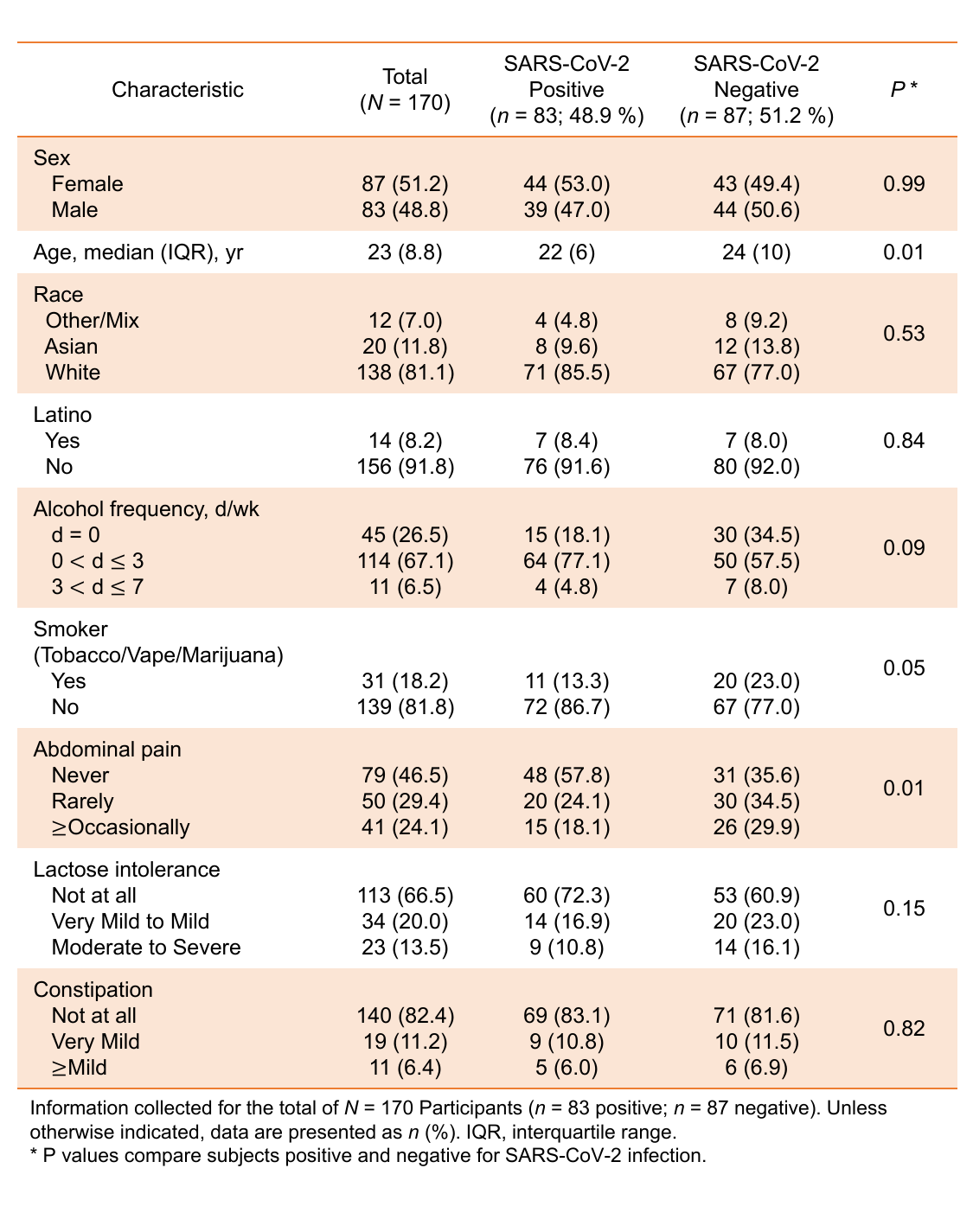}
  \captionof{table}{\textbf{Participant characteristics.}}
  \label{tab:Charac}
\endgroup

\newpage
\begingroup
  \centering
  \includegraphics[width=\linewidth]{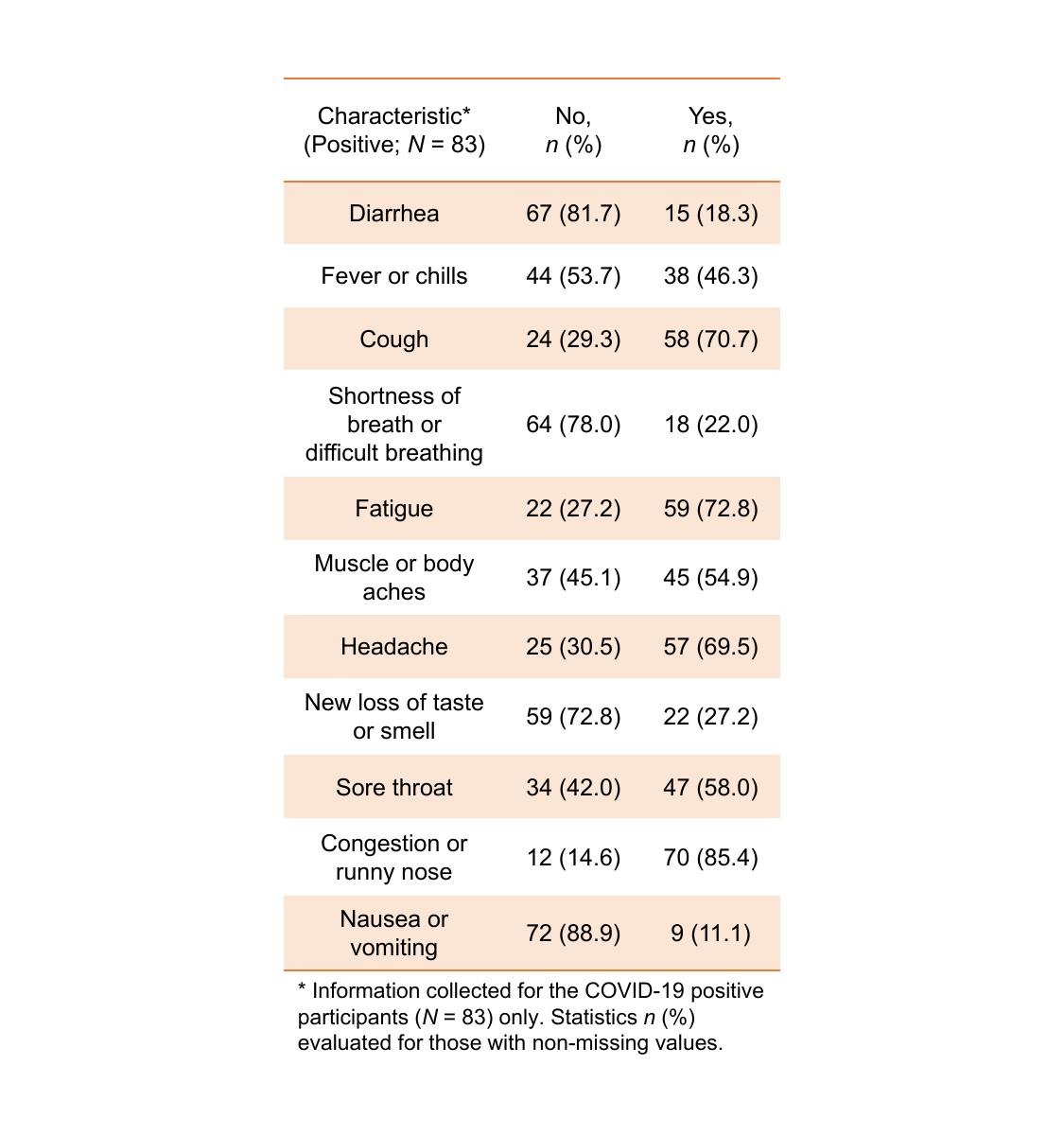}
  \captionof{table}{\textbf{COVID-19 symptoms experienced by the positive participants.}}
  \label{tab:Positive}
\endgroup

\clearpage
\newpage

\newpage

\newpage

\newpage

\clearpage
\newpage
\section*{Supplementary Information}
\renewcommand{\thefigure}{S\arabic{figure}}
\renewcommand{\thetable}{S\arabic{table}}
\setcounter{figure}{0}
\setcounter{table}{0}

\subsection*{Partial least squares-discriminant analysis (PLS-DA)} The principle of PLS regression and its usage for discriminant analysis, namely the PLS-DA algorithm, is briefly introduced here. The PLS regression toolbox used in our work was developed by MATLAB and implemented using the SIMPLS formulation. We discuss only the univariate response classification, corresponding to what is used in this work, but interested readers may consult Ref.~\cite{DEJONG_SIMPLS} for more details beyond this classification type and how the actual algorithm is implemented. We use bold upper case to denote matrices, bold lower case for vectors, and un-bold for scalars, with primes ($'$) denoting a matrix or vector transpose. Collected data used for the training process are represented by the $n \times p$ predictor variables matrix $\X_0$ and the $n \times 1$ univariate response variable vector $\y_0$. Here, $n$ is the total number of research subjects, $p$ is the total number of predictor variables. Both $\X_0$ and $\y_0$ are column-centered so that the covariance of different predictor variables with the response can be expressed by a $p \times 1$ column vector $\s_0 = \X_0'\y_0$. PLS regression relates $\X_0$ and $\y_0$ based on $\y_0 = \X_0 \b + \e$, where $\b$ is the $p \times 1$ coefficients estimate, $\X_0 \b$ is the explained component, and $\e$ is the fit residual. In contrast to least squares regression, where the coefficients estimate $\b$ is constructed by minimizing the residual sum of squares $\e'\e$, PLS regression constructs it based on the covariance $\s_0 = \X_0'\y_0$ to get more stabilized values of $\b$ and achieve more reliable predictive power. The formulation begins by projecting the predictor variables matrix $\X_0$ onto a new coordinate system $\T= \X_0 \R$ of reduced dimensionality spanned by a total of $A$ $(\leq p-1)$ PLS components, where $\R$ denotes the $p \times A$ weight transfer matrix and $\T$ denotes the $n \times A$ projected scores matrix. The construction of $\R$ is subject to two constraints: 1) the covariance vector $\T'\y_0$ is maximized for each entry, meaning each PLS component exhibits the largest possible covariance with the response; 2) the PLS components are orthonormal, i.e., columns of $\T$ satisfy $\t_i'\t_j = \delta_{ij}$ for any $i,j = 1,2,...,A$, where $\delta_{ij}$ is the Kronecker delta. The coefficients estimate $\b$ can be determined once $\R$ is known, since $\y_0=\T\T'\y_0=\X_0\R\R'\X_0'\y_0=\X_0\b$, and thus $\b=\R\R'\X_0'\y_0=\R\R'\s_0$. The process of determining $\R$ proceeds column by column. For the first iteration step $k = 1$, the maximization of the covariance of the first PLS component ($\t_k = \X_0 \r_k $) with the response, $\t_k'\y_0 = \r_k'\X_0'\y_0 = \r_k'\s_0 = \max$, constrains the first weight vector $\r_k$ ($k = 1$) to be along the direction of $\s_0$. For steps $k > 1$, the orthogonality condition, $\t_k'\t_i = \r_k'(\X_0'\t_i) = 0$ for $i=1,2,...,k-1$, requires the newly constructed $\r_k$ to be orthogonal to each of the $p \times 1$ vectors $\X_0'\t_i$ for $i = 1,2,...,k-1$. We define $\p_i \equiv \X_0'\t_i$ as the loading vectors. One may use the Gram-Schmidt process to find the orthonormal basis of the subspace $V_{k-1}$ spanned by the loading vectors $\p_i$ ($i = 1,2,...,k-1$) and then determine the $p \times p$ projection operator $\Proj$ for the orthogonal complement space $V^{\perp}_{k-1}$. This loosely constrains the direction of $\r_k$ to be within $V^{\perp}_{k-1}$, requiring $\r_k = \Proj \r_k$. Now, with the covariance maximization criteria, $\t_k'\y_0 = \r_k'(\Projt \s_0) = \max$, the direction of $\r_k$ is ultimately determined to be along the direction of the vector $\Projt \s_0$, which is the projection of the covariance vector $\s_0$ onto the subspace $V^{\perp}_{k-1}$. The iteration process proceeds until the directions of all $\r_k$ are determined, where the normalization condition $\T'\T = 1$ governs the magnitudes of $\r_k$. Finally, the coefficients estimate is determined and can be used for prediction of the response class for new observations based on $\y^{pred}_0 = \X^{new}_0 \b$, where the $m \times p$ matrix $\X^{new}_0$ is the testing data for a total of $m$ research subjects. The $m \times 1$ predicted values $\y^{pred}_0$ are translated proportionally into posterior probabilities and compared with a threshold value for response class assignment.

\clearpage
\newpage
\subsection*{Variable importance in the projection (VIP) scores} In PLS-DA, assessment of the importance of the predictor variables needs to consider 1) the weighting of a given predictor variable to form different PLS components and 2) the importance of different PLS components in explaining the response. Regarding 1), the formation of the $a$-th PLS component ($a = 1,2, ..., A$) takes the contribution from the $j$-th predictor variable with the normalized weight given by $w_{ja}/\|w_a\|$, where $w_{ja}$ is the $j$-th row $a$-th column element from the $p \times A$ weight matrix $\R$, and $\|w_a\| = (\sum_{j=1}^{p} w_{ja}^2)^{1/2}$ is the normalization. Regarding 2), we first note that the variance of the response among all observations $\y_0'\y_0$ is explained by the total of $A$ PLS components to the extent of $\yhat_0'\yhat_0$, where $\yhat_0 = \X_0 \b = \y_0 - \e$. The total percentage variance explained in the response, $(\yhat_0'\yhat_0/\y_0'\y_0) \times \SI{100}{\%}$, can be used for estimating the minimum number of PLS components needed for reliable predictions. The explained variance $\yhat_0'\yhat_0 = \yhat_0'\T\T'\yhat_0 = \sum_{a=1}^{A} (\yhat_0'\t_a)^2$ is further broken down into a summation of the square of the covariance of all PLS components with $\yhat_0$. We can thus evaluate the importance of the $a$-th PLS component by its variance explained $\q_a^2 \equiv (\yhat_0'\t_a)^2$, a quantity assigning larger importance to the PLS components that have larger covariance with the explained component, with the total variance explained by the $A$ PLS components given by $\sum_{a=1}^{A} \q_a^2$. Taking both 1) and 2) into account, the variable importance for the predictor variable $j$ summing over all the $A$ PLS components is proportional to $[\sum_{a=1}^{A} q_a^2 \cdot (w_{ja}/\|w_{a}\|)^2]^{1/2}$. From this, one can define its $\VIP$ score \cite{CHONG_VIP}, a metric for characterizing its importance, by:
\begin{equation}
    \VIP_j = \sqrt{\frac{p \cdot \sum_{a=1}^{A} [q_a^2 \cdot (w_{ja}/\|w_{a}\|)^2]}{\sum_{a=1}^{A} q_a^2}}
\end{equation}
Normalization ensures the mean square sums of the VIP scores among all predictor variables equals unity, $p^{-1} \sum_{j=1}^{p} \VIP_j^2 = 1$. Because of this normalization, predictor variables with $\VIP$ scores above (or below) unity can be regarded as important (or unimportant) variables.

\clearpage
\newpage
\subsection*{Variance explained by the PLS components.}  \label{totalVIP}
\begin{figure} [h]
\centering
\includegraphics[width=\textwidth]{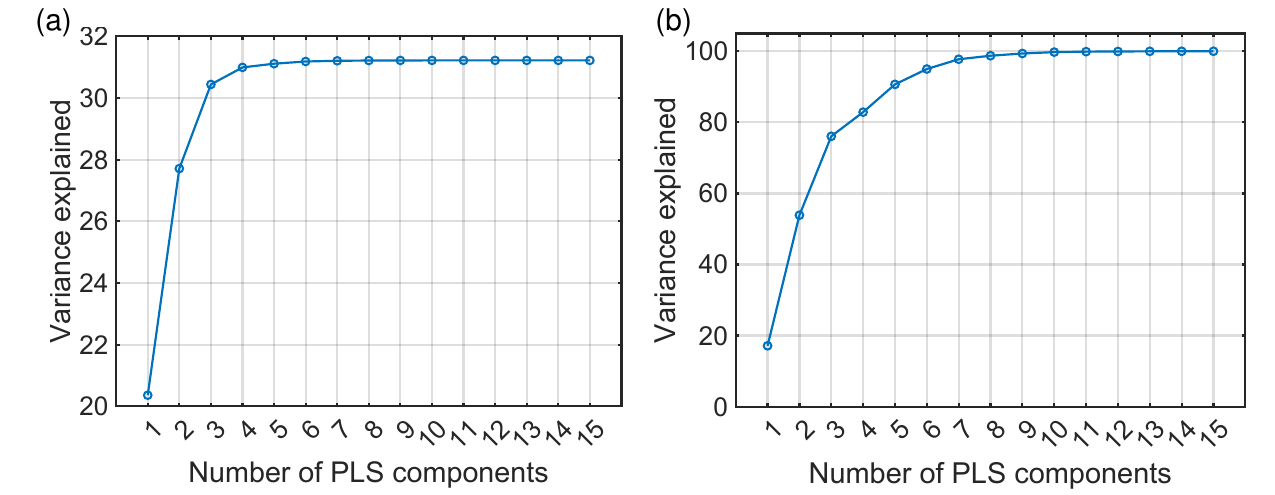}
\caption{\textbf{Total percentage variance explained in the response.} Results for (\a) the molecule-based approach and (\b) the pattern-based approach.}
\label{fig:VarExp}
\end{figure}

For SARS-CoV-2 infection classification, the total percentage variance explained in the response analyzed by the molecule-based and the pattern-based approaches for the complete data set (N = 170) are given in Figure~\ref{fig:VarExp}. We found a sharp rise in the variance explained for both the molecule-based and the pattern-based approaches when the number of PLS components constructed lies in the range from unity to five. A total of 15 PLS components were sufficient to saturate the percentage variance explained for both approaches. The lower variance explained obtained by the molecular species-based approach suggests fitting the spectroscopy data with more molecular species can better explain the response. 

\hspace{1cm}

\clearpage
\newpage
\subsection*{Averaging of the Receiver-Operating-Characteristics curves}
We performed averaging of the ROC curves using the non-parametric method adapted from Ref.~\cite{Chen_ROC_tilt}. This method ensured that: 1) the AUC of the averaged curve equaled the average AUC of individual cross-validation runs, and 2) the averaged AUC for a perfect (or random) classifier was equal to 1 (or 0.5). Proof for statement 1) can be found in the appendix of Ref.~\cite{Chen_ROC_tilt}, while statement 2) can be straightforwardly deduced from 1). In our work, we averaged the individual ROC curves vertically in the tilted space formed by rotating the ($\FP$,$\TP$) axes counter-clockwise by an angle $\theta < \pi/2$, where $\FP$ and $\TP$ denotes false positive rates and true positive rates, respectively. This enabled the averaging to be taken over singular functions. Any data point from an individual ROC curve could take its FP values from $\{(0, 1, 2, ... , \N)/\N\}$, and TP values from $\{(0, 1, 2, ... , \P)/\P\}$. Since we were using stratified sampling at the fixed testing set size $\Ltest = \P+\N$, different cross-validation runs preserved the total number of positives $\P$ and negatives $\N$. Hence, we chose $\theta = \arctan{(\P/\N)}$ such that the curve averaging in the tilted space would be performed to yield a total of $(\Ltest+1)$ sample points for plotting the averaged ROC curve. The $j$-th $(j = 0, 1, 2, ..., \Ltest)$ sample point represented the $j$-th observation in the testing set scanned over by the threshold line, and was obtained from the statistical mean over a total of the number of cross-validation runs of the $j$-th observation from each run.

\clearpage
\newpage
\subsection*{Uncertainty in the AUC} 
\begin{figure} [h]
\includegraphics[width=\textwidth]{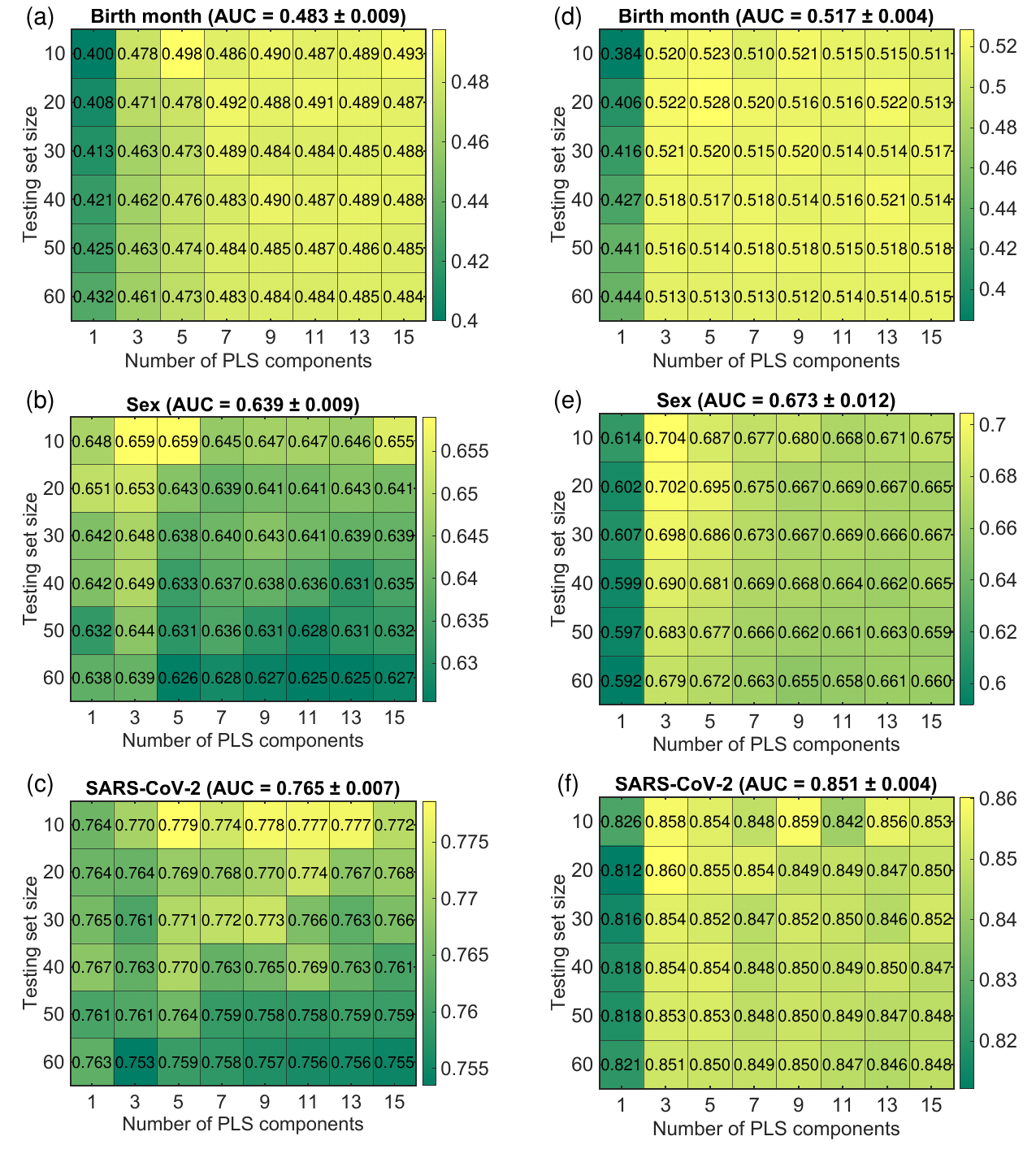}
\caption{\textbf{AUC calculated for different numbers of PLS components and different training and testing set partition ratios.} For different partition ratios, we show the testing set size in plotting the results. The training set size can be obtained by subtracting the testing set size from the complete data set size (N = 170). \a, \b, \c: results for the molecule-based approach, for birth month, sex, and SARS-CoV-2, respectively. \d, \e, \f: results for the pattern-based approach, for birth month, sex, and SARS-CoV-2, respectively.}
\label{fig:UncertaintyAUC}
\end{figure}

Uncertainty in the AUC for different response types was calculated using different numbers of PLS components and different partition ratios of the training and testing set (see Figure~\ref{fig:UncertaintyAUC}). For each number of PLS components and partition ratio used, an AUC value was calculated from the averaged ROC curve obtained from 1,000 cross-validation runs based on stratified random sampling. As seen in Figure~\ref{fig:UncertaintyAUC}, the AUC values calculated with only one PLS component were found to give worse prediction performance in general for both the molecule-based and the pattern-based approaches. This is understandable because both approaches showed limited total percentage variance explained when only one PLS component was constructed (see Figure~\ref{fig:VarExp}). For this reason, we calculated the mean and standard deviation of the AUC for each plot excluding those obtained using only one PLS component. Obtained values are reported in the title of each plot. The standard deviations were used as the uncertainty of AUC. The means were provided for reference. Note that in the main text the absolute values quoted for the AUC were computed using 15 PLS components, 140:30 training and testing partition ratio, and 10,000 cross-validation runs. We found the computed values using these settings matched the means obtained here to within the calculated uncertainty.

\clearpage
\newpage
\subsection*{Prediction performance summary.} 

A summary of binary response classification results for various response types is provided in Table~\ref{tab:Summary}. The obtained AUC shown for each response type were the mean and standard deviation calculated for the results obtained using 1,000 cross-validation runs based on stratified random sampling, evaluated at 3, 5, 7, ..., 15 PLS components, and at 10, 20, 30, ..., 60 test set size with training set size given by subtracting the testing set size from the complete data set. 

\newpage
\begingroup
  \centering
  \includegraphics[width=\linewidth]{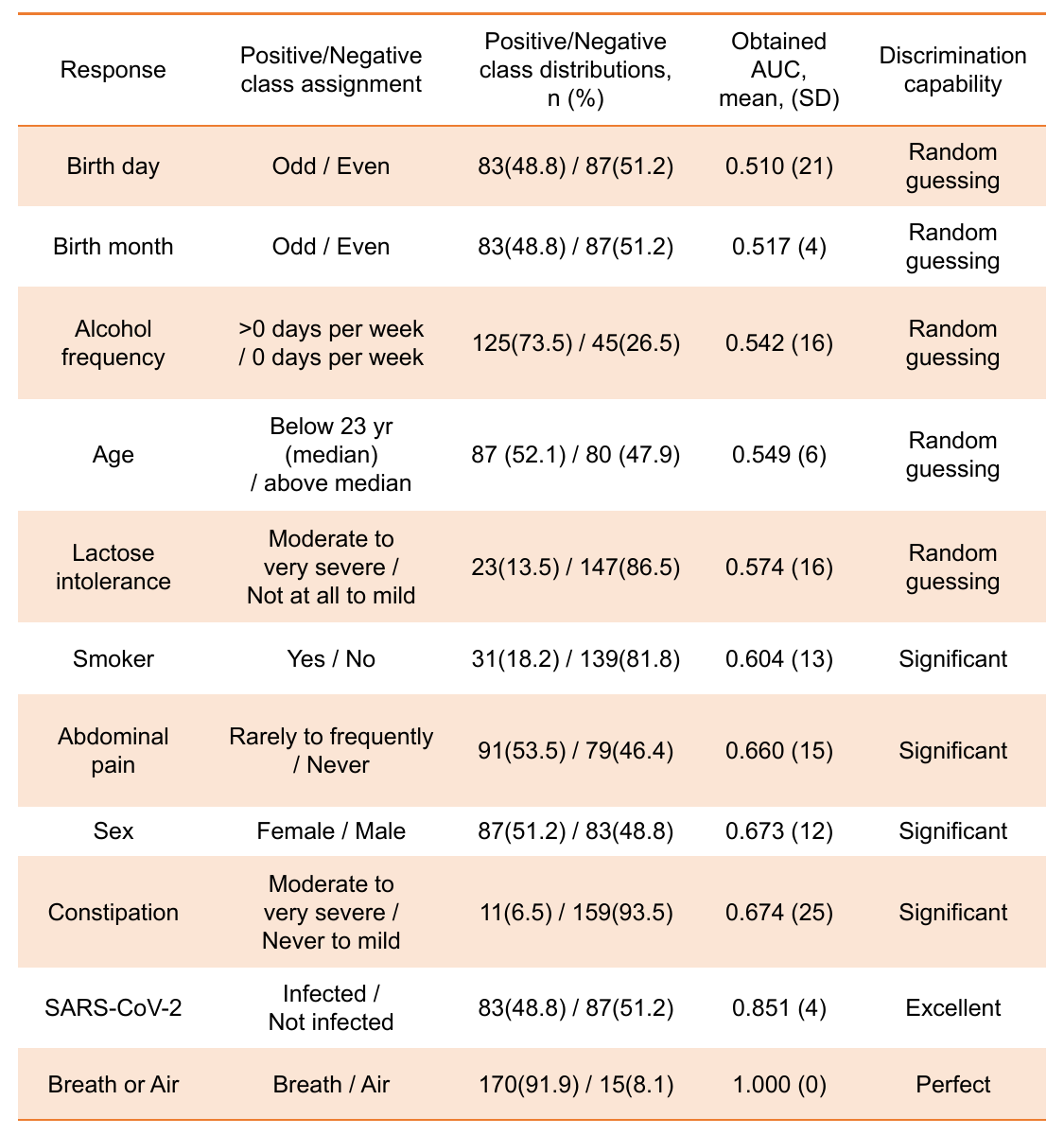}
  \captionof{table}{\textbf{Prediction performance summary.} AUC values were obtained for each response type and used for judgement of the discrimination capability.}
  \label{tab:Summary}
\endgroup

\newpage

\end{document}